\documentclass[12pt]{article}
\pdfoutput=1
\usepackage[margin=3cm]{geometry}
\usepackage[utf8]{inputenc}
\usepackage{amsmath}
\usepackage{amssymb}
\usepackage{authblk}
\usepackage{xcolor}
\usepackage{physics}
\usepackage{graphicx}
\usepackage{float}
\usepackage{tikz}
\usetikzlibrary{calc}
\usepackage{comment}
\usepackage{acronym}
\newacro{EFT}[EFT]{effective field theory}
\newacro{AdS}[AdS]{anti-de Sitter}
\newacro{dS}[dS]{de Sitter}
\acrodefplural{EFT}{effective field theories}
\newacro{SUSY}[SUSY]{supersymmetry}
\newacro{KK}[KK]{Kaluza-Klein}

\usepackage{hyperref}
\usepackage{varioref}       
\usepackage{cleveref}   
\usepackage{xcolor}
\crefname{table}{table}{tables}
\Crefname{table}{Table}{Tables}
\crefname{figure}{figure}{figures}
\Crefname{figure}{Figure}{Figures}

\definecolor{tealblue}{rgb}{0.21, 0.56, 0.63}
\hypersetup{colorlinks=true,allcolors = tealblue,linktocpage=true}

\usepackage[
        backend=bibtex,
        sorting=none,
        style=phys,
        eprint=true,
        doi=false,
        biblabel=brackets
        ]{biblatex}

\bibliography{bibliography}

\newenvironment{eqaed}
    {\begin{equation}
    \begin{aligned}
    }
    { 
    \end{aligned}
    \end{equation}
    \ignorespacesafterend
    }

\newcommand{\nc}{\newcommand}
\nc{\lb}{\llbracket}
\nc{\rb}{\rrbracket}
\nc{\gl}{\llbracket}
\nc{\gr}{\rrbracket}
\nc{\del}{\partial}
\nc{\tri}{\hspace{-3.5pt}\vartriangle\hspace{-3.5pt}}
\nc{\blacktri}{\blacktriangle}

\nc{\eq}[1]{\begin{equation}
                     \begin{split} #1 \end{split}
                     \end{equation}}
\nc{\ul}{\underline}
\nc{\ov}{\overline}

\nc{\fa}{\hat}
\nc{\fb}{\MakeUppercase}
\nc{\fc}{\tilde}
\nc{\Lie}{{\cal L}} 
\nc{\lambdabar}{{\mkern0.75mu\mathchar '26\mkern -9.75mu\lambda}}

\allowdisplaybreaks[2]
\numberwithin{equation}{section}

\DeclareUnicodeCharacter{2212}{-}
\begin{document}
	\begin{flushright}
		MPP-2026-92
	\end{flushright}

\vspace{0.5cm}
\begin{center}
  {\LARGE \bf Quintessential
$\alpha$-attractors fit DESI
} 
\vspace{0.2cm}

\end{center}

\vspace{0.15cm}
\begin{center}
Alessandro Borys$^{1,2}$,
Joaquin Masias$^3$,
Marco Scalisi$^{1,2}$ \\[0.2cm]
\end{center}

\vspace{0.0cm}
\begin{center} 
{\footnotesize
\vspace{0.25cm} 
\emph{$^1$Department of Physics, University of Catania, \\
Via Santa Sofia 64, 1-95125 Catania, Italy}\\[0.1cm]
\vspace{0.25cm} 

\emph{$^2$INFN-Sezione di Catania, \\
Via Santa Sofia 64, 1-95123 Catania, Italy}
\\[0.1cm]
\vspace{0.25cm} 

\emph{$^3$Max-Planck-Institut f\"ur Physik (Werner-Heisenberg-Institut), \\ 
Boltzmannstr. 8, 85748 Garching, Germany}\\[0.1cm]
\vspace{0.25cm} 
}
\end{center} 

\vspace{0.3cm}

\begin{abstract}
\noindent
We study quintessence in $\alpha$-attractor models in light of recent DESI indications for dynamical dark energy. We show that the \emph{knee} of the attractor potential provides an excellent approximation to the axion-like quintessence model used as a DESI benchmark. This leads to a simple relation between the axion decay constant $f_a$ and the attractor parameter $\alpha$, allowing the experimental constraints to be translated into a preference for $\alpha=\mathcal{O}(1)$, in agreement with string-motivated expectations.
We solve the background dynamics numerically and find good agreement with the DESI-preferred evolution of $w(z)$ up to $z\sim\mathcal{O}(1)$. More generally, we point out that the agreement between axion-like and attractor potentials reflects a common requirement imposed by the data: today’s potential energy and slope are both of order the Hubble scale in Planck units. We finally comment on the origin of the required initial conditions, which can naturally arise in multifield attractor scenarios.

\end{abstract}

\newpage

\section{Introduction}

The nature of dark energy remains one of the central open problems in modern cosmology \cite{Weinberg:1988cp}. The simplest and most economical explanation is provided by a cosmological constant, yet recent data releases from the Dark Energy Spectroscopic Instrument (DESI) \cite{DESI:2025zgx}  have opened the possibility that the dark energy sector may be dynamical. When combined with cosmic microwave background (CMB) and supernova measurements \cite{Brout:2022vxf, Rubin:2023ovl, DES:2024jxu, Sabogal:2025jbo}, DESI baryon acoustic oscillation data show a preference for an evolving equation of state, motivating a closer examination of simple and theoretically well motivated models of quintessence \cite{Ratra:1987rm,Caldwell:1997ii,Tsujikawa:2013fta,Peebles:1987ek, Wetterich:1994bg} .

Scalar field models provide a natural framework to describe both early- and late-time accelerated expansion \cite{Baumann:2009ds,Martin:2013tda,Tsujikawa:2013fta}. In particular, $\alpha$-attractor models \cite{Kallosh_2013,Roest:2015qya,Scalisi:2015qga,Scalisi:2018eaz, Kallosh:2022feu, Kallosh:2025jsb, Kallosh:2025ijd} have emerged as a robust and predictive class of inflationary scenarios, characterized by universal predictions that are largely insensitive to the microscopic details of the scalar potential \cite{Kallosh:2015zsa,Carrasco:2015uma,Kallosh:2013yoa}. This universality originates from the hyperbolic geometry of the scalar field space, which stretches the potential near the boundary of moduli space and leads to extended plateau regions \cite{Carrasco:2015uma, Kallosh:2015zsa}.

\begin{figure}
    \centering
\includegraphics[width=0.55\linewidth]{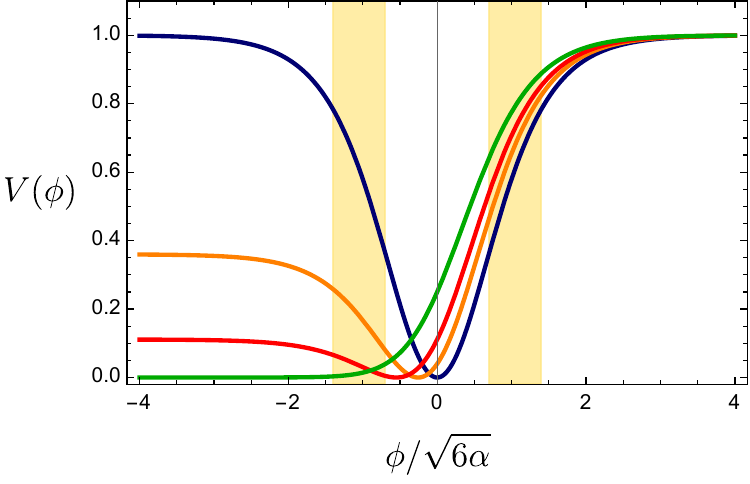}
    \caption{\textit{Example of potential with attractor value $\alpha$, for $c=0,\frac{1}{4}, \frac{1}{2}$ and $1$, normalized to 1 at infinity.. The highlighted region will be the relevant region for quintessence matching DESI results.}}
    \label{fig:potentialplot}
\end{figure}

$\alpha$-attractors have also been explored in the context of quintessence \cite{Dimopoulos:2017zvq,Alestas:2024eic, Scalisi:2015qga} (for broader reviews of string cosmology and dark energy in string theory see \cite{Cicoli:2023opf, Andriot:2026lac}). Existing studies have predominantly focused on the asymptotic regimes of the potential, namely the de Sitter plateau or the exponentially decaying tail.  In this work, we instead focus on a different region: the intermediate field range, or \emph{knee}, where the potential interpolates between these two asymptotic behaviors. This region is shown schematically in Fig. \ref{fig:potentialplot}.
This regime is particularly relevant for late-time cosmology, as it naturally allows for a thawing evolution in which the scalar field remains frozen for most of cosmic history and only begins to roll down along a concave potential near the present epoch, in line with the behavior suggested by recent DESI data \cite{DESI:2025fii}. Similar thawing regimes have been found also in other contexts (see e.g. \cite{Shlivko:2024llw,Anchordoqui:2025fgz, Anchordoqui:2025epz, Shiu:2026edl,Shlivko:2025fgv, Urena-Lopez:2025rad}).

Motivated by recent developments, we investigate whether quintessential $\alpha$-attractor potentials can reproduce the preferred behavior of the dark energy equation of state. We show that, in the field range relevant for quintessence, the attractor potential provides an excellent approximation to the axion-like model used as a benchmark in DESI analyses \cite{DESI:2025fii}, leading to a simple relation between the effective axion decay constant and the attractor parameter $\alpha$. This correspondence allows the DESI constraints to be translated into a preference for $\alpha=\mathcal{O}(1)$, in agreement with expectations from string-motivated constructions. We further solve the cosmological dynamics numerically and find that the resulting evolution of the equation of state $w(z)$ agrees well with the DESI-preferred region up to redshift $z \sim \mathcal{O}(1)$.

Our analysis suggests also that the apparent success of both axion-like and attractor potentials can be traced back to a common requirement imposed by the data: the scalar potential today must have both its value and its slope of order the Hubble scale in Planck units. This points toward a regime in which dark energy is not exactly frozen, but is already evolving away from a cosmological constant. Within this perspective, the knee of the attractor potential emerges as a particularly natural region in which to realize such dynamics.

The paper is organized as follows. In Sec. \ref{sec:stringsourcedalpha_tech} we review the origin of $\alpha$-attractors from hyperbolic moduli spaces in supergravity and string theory. In Sec. \ref{section3} we compare the attractor potential to the axion-like benchmark used in DESI analyses and study the resulting cosmological evolution. In Sec. \ref{Double} we argue why both attractor and axionic potentials are able to reproduce successfully DESI data. Finally, in Sec. \ref{choice} we discuss the origin of the required initial conditions in a simple multi-field setup. 

\section{Hyperbolic spaces and attractor potentials}
\label{sec:stringsourcedalpha_tech}

The scalar field appearing in the attractor potential can be identified with the real part of a complex modulus parametrizing a hyperbolic scalar manifold of constant negative curvature. The relevant moduli space is
\begin{equation}
\mathcal{M} \simeq \frac{SL(2,\mathbb{R})}{SO(2)} \simeq \mathbb{H}\,,
\end{equation}
which admits equivalent descriptions in Poincaré upper half-plane ($T, \bar T$), or in Poincaré disk variables ($Z, \bar Z$), related by
\begin{equation}
\label{cayley}
Z=\frac{T-1}{T+1}\,,
\qquad
T=\frac{1+Z}{1-Z}\,.
\end{equation}

In disk coordinates, the $\mathcal{N}=1$ SUGRA K\"ahler potential reads
\begin{equation}
K=-3\alpha\log(1-Z\bar Z)\,,
\end{equation}
yielding the K\"ahler metric
\begin{equation}
ds^2=\frac{3\alpha}{(1-Z\bar Z)^2}\,dZ\,d\bar Z\,,
\end{equation}
with constant negative curvature
\begin{eqaed}
    \mathcal{R_K}=-\frac{2}{3\alpha}\,,
\end{eqaed}
depending directly on the attractor parameter $\alpha$. 
Restricting to the real disk direction at vanishing axion, $Z=\bar Z\equiv \xi$, the Lagrangian takes the form
\begin{equation}
\frac{1}{\sqrt{-g}}\mathcal{L}
=
\frac{R}{2}
-
\frac{3\alpha\,(\partial_\mu \xi)^2}{(1-\xi^2)^2}
-
V(\xi)\,.
\end{equation}
The canonically normalized field is then obtained through
\begin{equation}
\xi=\tanh\left(\frac{\phi}{\sqrt{6\alpha}}\right),
\end{equation}
which gives
\begin{equation}
\frac{1}{\sqrt{-g}}\mathcal{L}
=
\frac{R}{2}
-
\frac{(\partial_\mu \phi)^2}{2}
-
V\left(
\tanh\frac{\phi}{\sqrt{6\alpha}}
\right).
\end{equation}
In this parametrization, the boundary of moduli space corresponds to $|\xi|\longrightarrow 1$, which is reached only at infinite distance $\phi\longrightarrow\pm\infty$ \cite{Scalisi:2018eaz,vandeHeisteeg:2023uxj}. Consider for instance a quadratic scalar potential of the form
\begin{equation}
V(\xi)\simeq(\xi-c)^2.
\end{equation}
with $0\leq c\leq1$  a constant that governs how the potential approaches the asymptotic regions. Expressed in terms of the canonically normalized field $\phi$, this becomes
\begin{equation}\label{eq:V}
V(\phi)=
V_1
\left(
\tanh\frac{\phi}{\sqrt{6\alpha}}
-c
\right)^2,
\end{equation}
which is the so-called T-model potential. The asymptotic plateau behavior therefore does not arise from imposing an exponentially flat potential by hand, but rather from the hyperbolic geometry of field space itself. Any sufficiently regular potential written in disk variables becomes asymptotically flat near the boundary once expressed in terms of the canonically normalized field.

Hyperbolic moduli spaces of this type are ubiquitous both in supergravity and in string compactifications. For example, M-theory compactified on $T^7$ yields $d=4$, $\mathcal N=8$ maximal SUGRA \cite{Cremmer:1978ds,Cremmer:1979up,deWit:1982bul} which allows for a consistent truncation down to $\mathcal N=1$. The full scalar manifold contains a submanifold that survives the truncation,
\begin{equation}
\mathcal M_{\rm scalar}\simeq \frac{E_{7(7)}}{SU(8)}\supset \left(\frac{SL(2,\mathbb{R})}{SO(2)}\right)^7\,,
\end{equation}
namely seven copies of the hyperbolic half-plane, with Kähler potential
\begin{equation}
K = -\sum_{i=1}^7 \log\!\big(T_i+\bar T_i\big),
\end{equation}
often referred to as the seven disk manifold \cite{Ferrara:2016fwe}. A single effective $\alpha$-attractor can be obtained by aligning $n$ of the moduli,
\begin{equation}
T_1=\cdots=T_n \equiv T,
\qquad
T_{n+1},\dots,T_7=\text{const.},
\end{equation}
so that $n$ directions are isotropic while $7-n$ directions remain fixed. In this limit the Kähler potential is given by
\begin{equation}
K = -3\alpha\log\!\big(T+\bar T\big),
\qquad
\alpha=\frac{n}{3},\qquad n=1,\dots,7.
\end{equation}
In this way, the allowed attractor values are upper bounded by the maximum number of internal directions that can simultaneously decompactify.\footnote{The same coset manifold can be obtained from M-theory on a 7-manifold with $G_2$ holonomy and Betti numbers $(b_0, b_1, b_2, b_3, b_4, b_5, b_6, b_7)=(1,0,0,7,7,0,0,1)$. This is known as the $\mathcal{N}=1$,  $d=4$ \textit{curious}  supergravity \cite{Duff:2010vy}.}

Similar structures also appear in Type IIB Calabi--Yau orientifold compactifications with ISD fluxes \cite{Giddings:2001yu, Kachru:2003aw, Balasubramanian:2005zx}, where only the Kähler moduli remain massless. The relevant tree-level Kähler potential is then
\begin{equation}
K \,=\, -2\log \mathcal V,
\label{eq:K_IIB_Kahler}
\end{equation}
where $\mathcal V=\mathcal V(T^a+\bar T^a)$ is the total internal volume and $T^a=\tau^a+i a^a$ are the  complexified 4-cycle volumes. In the limit where a single cycle becomes large, the volume typically scales as
\begin{equation}
\mathcal{V}\propto \tau^{3\alpha/2},
\end{equation}
which implies
\begin{equation}
K\simeq -3\alpha\log(T+\bar T),
\end{equation}
with discrete values
\begin{equation}
3\alpha=1,2,3,
\end{equation}
depending on whether the dominant contribution arises from a 2-, 4-, or 6-cycle becoming large \cite{Grimm:2004uq}.  In all of these constructions, the attractor parameter $\alpha$ is therefore not arbitrary, but is instead fixed by the geometry of the compactification manifold and naturally takes $\mathcal{O}(1)$ rational values.

\section{\texorpdfstring{$\alpha$-attractors and quintessence}{alpha-attractors and quintessence}}
\label{section3}

We now examine whether $\alpha$-attractor potentials are compatible with the DESI constraints on late-time cosmic acceleration. Using the Chevallier--Polarski--Linder (CPL) parameterization~\cite{Chevallier:2000qy,Linder:2002et}, the dark energy equation of state is written as a function of the scale factor $a$ (or equivalently redshift $z$) as
\begin{equation}
    w(z)=w_0+w_a(1-a)=w_0+w_a\frac{z}{1+z}.
\end{equation}

By combining BAO, CMB, and supernova data, the DESI collaboration identified a clear preference for a region in the $(w_0, w_a)$ parameter space where $w_0+w_a<-1$, with results depending on the Supernovae dataset used:

\begin{center}
\renewcommand{\arraystretch}{1.15}
\begin{tabular}{lcc}
\hline
\rule{0pt}{2.5ex} Dataset & $w_0$ & $w_a$ \\[2pt]
\hline
\textcolor{gray}{DESI+CMB+PantheonPlus} & $-0.838 \pm 0.055$ & $-0.62 \pm 0.21$ \\
\textcolor{gray}{DESI+CMB+Union3}       & $-0.667 \pm 0.088$ & $-1.09 \pm 0.29$ \\
\textcolor{gray}{DESI+CMB+DESY5}        & $-0.752 \pm 0.057$ & $-0.86 \pm 0.22$ \\
\hline
\end{tabular}
\end{center}

At face value, this behaviour might suggest a transition in $w(z)$ from a regime violating the null-energy condition at higher redshift to one consistent with it at late times. However, such an interpretation is not necessary. Simple thawing quintessence models~\cite{Caldwell:2005tm} can naturally populate the same region of parameter space preferred by the data~\cite{Shlivko:2024llw,Anchordoqui:2025fgz,Anchordoqui:2025epz, Shlivko:2025fgv, Urena-Lopez:2025rad}.

Since cosmological observables are predominantly sensitive to low redshift and uncertainties increase significantly at higher redshift, the CPL parametrization should be interpreted as a local expansion around the present epoch ($a=1$). In this sense, $w_a$ captures the instantaneous derivative of the equation of state rather than a global slope. Extrapolating the linear CPL form to high redshift implicitly assumes the suppression of higher-order terms, which is not generically justified.

We now compare the $\alpha$-attractor potential \eqref{eq:V} with the thawing quintessence model used in the DESI analysis, based on the axion-like potential
\begin{equation}\label{eq:Va}
    V_a = m_a^2 f_a^2 \left(1 + \cos\frac{\phi}{f_a}\right).
\end{equation}
Here $m_a$ and $f_a$ denote respectively the mass and decay constant of the axion field. This potential is used as a representative proxy to identify regions of parameter space compatible with the data (for recent analyses in light of DESI, see e.g.~\cite{Anchordoqui:2025epz,Shiu:2026edl}).
DESI data suggest that the scalar field evolves in a hilltop-like regime, rolling down the concave region of the potential, with initial values in the range $\phi/f_a \in [0.7,1.0]$ and values today in $\phi/f_a \in [1.1,1.4]$. This corresponds to a sub-Planckian field excursion $\Delta\phi \sim 0.4\, f_a$~\cite{DESI:2025zgx}.

\subsection{Least-squares fit} 

In this section, we perform a numerical fit of \eqref{eq:V} against \eqref{eq:Va} in the relevant field range for quintessence as given by DESI~\cite{DESI:2025zgx}. We parameterize the potentials as
\begin{align}
    V_\alpha(\phi)&=V_1 \left[\tanh\left({\dfrac{\phi}{\sqrt{6\alpha}}}-\Delta\right)-c\right]^2, \label{eq:Va3} \\
    V_a(\phi)&=\frac{1}{2}V_0\left(1+\cos \frac{\phi}{f_a}\right), \label{eq:Va2}
\end{align}
where $V_1$ and $V_0$ are normalization constants of the order of the current dark energy density scale, $\rho_0\simeq 10^{-120} M_{\rm P}^4$, and $\Delta$ is a shift encoding the location at which the field starts rolling.

\begin{figure}
\begin{tabular}{cc}
        \centering
    \includegraphics[width=0.455\linewidth]{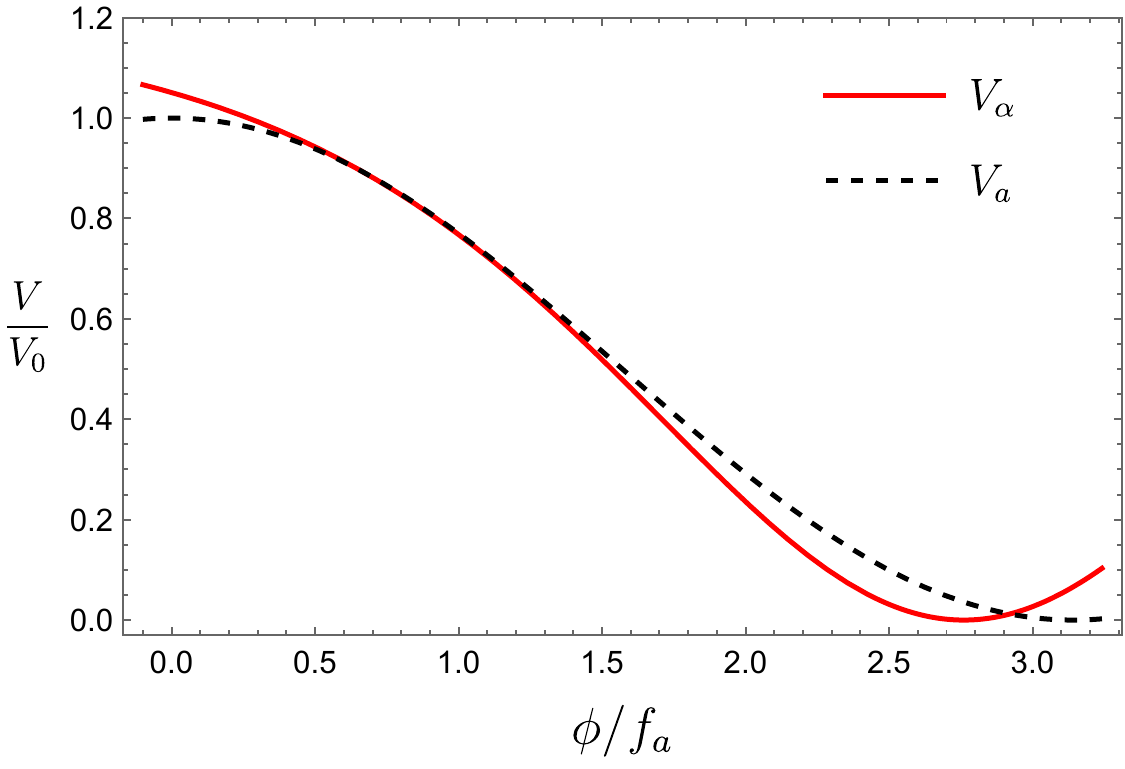} &    \centering
    \includegraphics[width=0.545\linewidth]{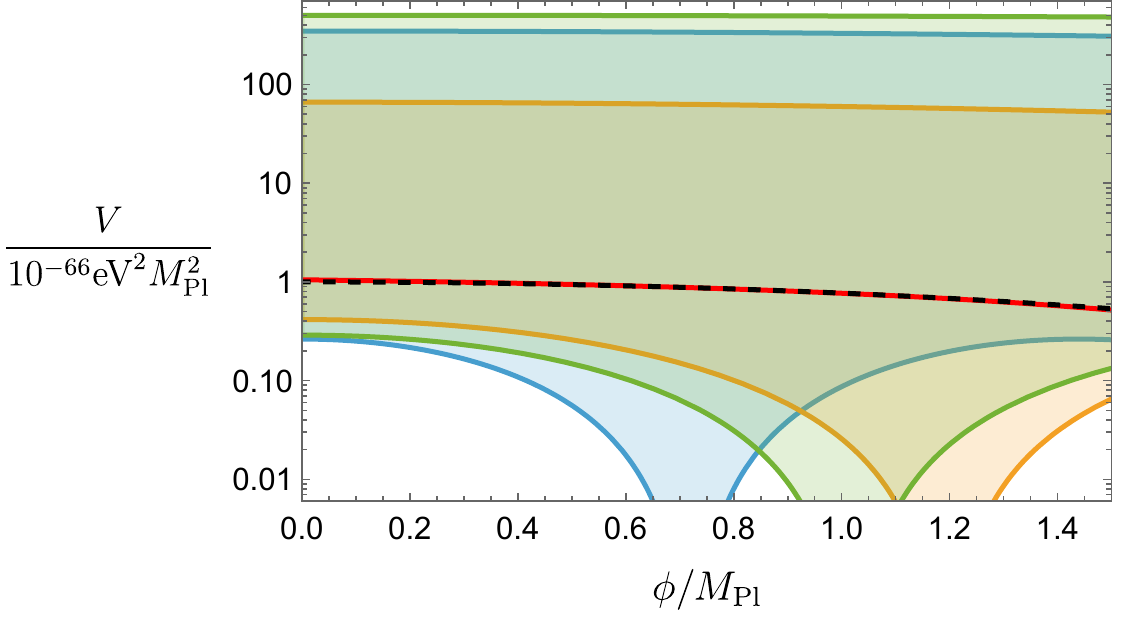} 
\end{tabular}
    \caption{\textit{Left: fit of the attractor potential with $c=0$ to the axion potential of \cite{DESI:2025zgx}. This is independent on the normalization or the value of the action decay constant of the axion potential. Right: results of the fit compared to the experimental bounds on the axion potential. The fit shown corresponds to $\alpha=1/3$, with $V_1=1.19\,V_0$, $\Delta=1.74$, and $f_a=\sqrt{\alpha/0.42}$.}}
    \label{fig:fitresults}
\end{figure}

For each value of $f_a$, we sample the axion potential $V_a(\phi)$ on a uniform grid in the interval $\phi/f_a \in [0.7,1.1]$, and denote the sampled points by $\phi_i$. The parameters $(V_1,\alpha,\Delta)$ of the attractor potential are then obtained by minimizing the function
\begin{equation}
\chi^2(V_1,\alpha,\Delta)=\sum_{i=1}^{N}\left[V_\alpha(\phi_i;V_1,\alpha,\Delta)-V_a(\phi_i;V_0,f_a)\right]^2.
\end{equation}
The overall normalization of the axion potential is fixed to $V_0=1$, while $V_1$, $\alpha$, and $\Delta$ are treated as free parameters. Repeating the procedure for $0.1\leq f_a\leq 4$ and for fixed $c$, one obtains an empirical relation between $f_a$ and $\alpha$. An example of the fit for $c=0$ is shown in Fig. \ref{fig:fitresults}.

\begin{figure}
\centering
\includegraphics[width=0.9\textwidth]{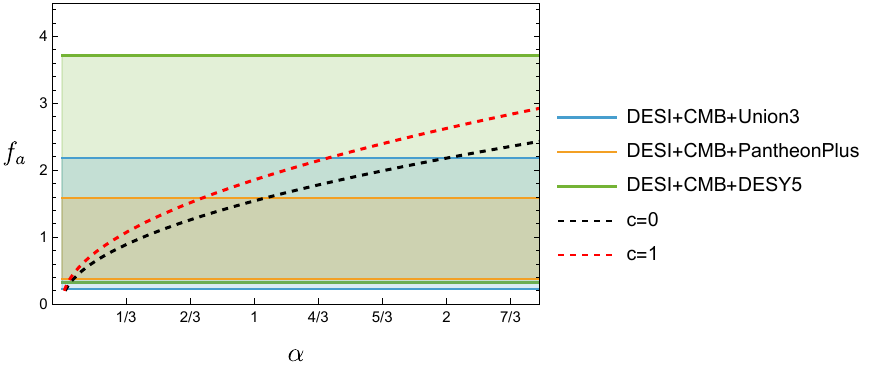}
\caption{\textit{
Fit of $f_a$ as a function of $\alpha$ for the attractor potential \eqref{eq:V} with $c=0$ and $c=1$. The shaded bands indicate the DESI-allowed range for $f_a$.
}}
\label{fig:redd}
\end{figure}

The fit yields a simple relation between the effective decay constant and the attractor parameter. In particular, one finds $f_a=\sqrt{\alpha/0.42}$ for $c=0$, and $f_a=\sqrt{\alpha/0.29}$ for $c=1$. Repeating the procedure for several values of $c$ in the interval $[0,1]$, we obtain the empirical relation\footnote{
One can also derive an analytic relation between the two potentials by matching the potential and its first two derivatives around $\phi = \epsilon\, f_a$. This leads to the relation $f_a = \sqrt{\alpha}\, g(\epsilon, c)$, with 
\begin{equation*}
    g(\epsilon, c) = \frac{1}{2} \sqrt{\frac{3}{2}}
    \tan \left(\frac{\epsilon}{2}\right)
    \left(
    \sqrt{c^2-1+\csc ^4\left(\frac{\epsilon}{2}\right)}+c
    \right).
\end{equation*}
For $\epsilon \in [0.7,1.1]$, this reproduces the numerical fit in \eqref{falpha} for $c \in [0,1]$.
}
\begin{equation}\label{falpha}
f_a=\frac{\sqrt{\alpha}}{\sqrt{0.42 - 0.13\, c}}\,.
\end{equation}
This resulting relation is shown in Fig.~\ref{fig:redd}, where the cases $c=0$ and $c=1$ are displayed together with the DESI-allowed region for $f_a$.

The DESI bounds on the axion decay constant $f_a$ (in Planck units) and on the axion mass $m_a$ (in units of $10^{-33}\,$eV), obtained from BAO, CMB, and the three supernova compilations, are summarized below:

\begin{center}
\renewcommand{\arraystretch}{1.15}
\begin{tabular}{lcc}
\hline
\rule{0pt}{2.5ex} Dataset & $f_a$ & $m_a$ \\[2pt]
\hline
\textcolor{gray}{BAO+CMB+PantheonPlus} & $0.38 < f_a < 1.59$ & $1.20 < m_a < 3.63$ \\
\textcolor{gray}{BAO+CMB+Union3}       & $0.23 < f_a < 2.19$ & $1.59 < m_a < 6.03$ \\
\textcolor{gray}{BAO+CMB+DESY5}        & $0.32 < f_a < 3.72$ & $1.18 < m_a < 4.27$ \\
\hline
\end{tabular}
\end{center}

Using the relation \eqref{falpha}, these constraints can be translated into bounds on the attractor parameter $\alpha$. This allows one to reinterpret the DESI-preferred region for axion quintessence directly in terms of attractor quintessence models:

\begin{center}
\renewcommand{\arraystretch}{1.15}
\begin{tabular}{lcc}
\hline
\rule{0pt}{2.5ex} Dataset & $c=0$ & $c=1$ \\[2pt]
\hline
\textcolor{gray}{BAO+CMB+PantheonPlus} & $0.06 < \alpha < 1.06$ & $0.04 < \alpha < 0.73$ \\
\textcolor{gray}{BAO+CMB+Union3}       & $0.02 < \alpha < 2.01$ & $0.02 < \alpha < 1.39$ \\
\textcolor{gray}{BAO+CMB+DESY5}        & $0.04 < \alpha < 5.80$ & $0.03 < \alpha < 4.00$ \\
\hline
\end{tabular}
\end{center}

Interestingly, the preferred values are generically of order unity, $\alpha=\mathcal{O}(1)$, in agreement with the values typically encountered in string theory and M-theory constructions. Once the mapping between $f_a$ and $\alpha$ is established, the observed range of the axion mass $m_a$ determines the overall normalization scale of the attractor potential.

These results suggest that attractor quintessence models with $\alpha=\mathcal{O}(1)$ provide a viable effective description of the DESI-preferred axion quintessence region.

\subsection{\texorpdfstring{Numerical computation of $w(z)$}{Numerical computation of w(z)}}

We now compute the equation of state $w(z)$ for the attractor potential \eqref{eq:V} directly. The relevant equations are the equation of motion for the scalar field $\phi$ and the Friedmann equation,
\begin{equation}
\ddot{\phi} = -3H\dot{\phi} - V'(\phi)\,,
\end{equation}
\begin{equation}
\dot{a} = a\,H= \dfrac{a}{\sqrt{3}}\sqrt{\dfrac{1}{2}\dot\phi^2+V(\phi)+\dfrac{\rho_{m,0}}{a^3}}\,,
\end{equation}
where the dot and the prime denote derivatives with respect to cosmic time and to $\phi$, respectively. We impose the initial conditions
\begin{equation}
w_\phi(0) = w_0\,,\quad \Omega_\phi(0)=\Omega_{\phi,0} = 0.69\,, \quad a(0) = 1, \quad H(0)=H_0\,,
\end{equation}
with the subscript $0$ denoting present-day quantities. The equation of state and the dark energy fraction read
\begin{eqaed}
    w_\phi=\dfrac{P_\phi}{\rho_\phi}=\dfrac{\frac{1}{2}\dot\phi^2-V(\phi)}{\frac{1}{2}\dot\phi^2+V(\phi)}\,,
    \qquad
    \Omega_{\phi}=\dfrac{\rho_{\phi}}{3H^2}\,.
    \label{eq:wphi}
\end{eqaed}
Combining these with the initial conditions one obtains
\begin{eqaed}
    \dfrac{\dot\phi^2_0}{2}=\dfrac{3}{2}\Omega_{\phi,0}(1\,+\,w_{0})H_0^2\quad     \text{and} \quad V(\phi_0)=\dfrac{3}{2}\Omega_{\phi,0}(1\,-\,w_{0})H_0^2\,,
    \label{eq:phidotsqandv}
\end{eqaed}
thus fixing the initial conditions for the numerical evolution. We then choose a present-day value of $w_0$ in agreement with experimental data. As expected from the direct comparison to the axion potential, the attractor model can fit experimental results for redshift up  to $z\simeq \mathcal{O}(1)$.
\begin{figure}
    \centering
        \begin{tabular}{cc}
              \includegraphics[width=0.64\linewidth]{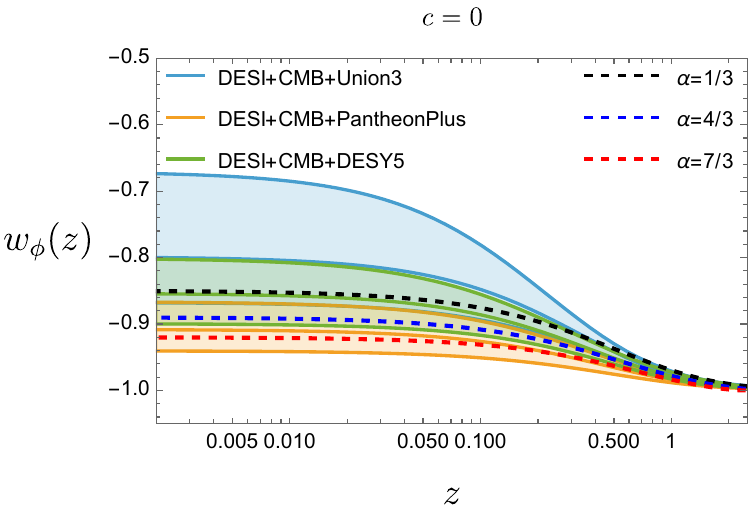}     \\
        \end{tabular}

    \caption{\textit{
Equation of state for the attractor potential \eqref{eq:V} with $c=0$ as a function of redshift, superimposed on the DESI constraints, for different values of $\alpha$. The initial conditions are chosen such that $w_0=-0.85,\,-0.89,\,-0.92$ for $\alpha=1/3,\,4/3,\,7/3$, respectively. The corresponding plot for $c=1$ is nearly indistinguishable.
}}
    \label{fig:weos}
\end{figure}
This is shown in Fig. \ref{fig:weos}, where we consider some initial conditions for the fields as they roll down the potential. We stress that the fields start rolling in the decaying (or \textit{knee}) part of the potential, rather than on the plateau.

\section{DESI constraints on $V$, $V'$ and $V''$}\label{Double}
We now turn to the question of why both axion and attractor potentials can fit the recent DESI data on dynamical dark energy. We can compute $w_a$ as
\begin{eqaed}
    w_{a}=-\dfrac{dw_{\phi}}{da}\bigg\vert_{a=1}=-\dfrac{1}{H_0}\dfrac{dw_{\phi}}{dt}\bigg\vert_{t=0}\,,
\end{eqaed}
where $w_a$ is here interpreted as an instantaneous quantity, namely as the first term in a Taylor expansion of $w_\phi$ around $a=1$, rather than as the global slope of the CPL parametrization.

Using \eqref{eq:wphi}, \eqref{eq:phidotsqandv}, and the equation of motion for $\phi$, one finds
\begin{eqaed}
    w_a=3\left(1-w_0^2\right) +2\sqrt{\dfrac{w_0+1}{3\,\Omega_{\phi,0}}}\dfrac{V'(\phi_0)}{H_0^2}\,.
\end{eqaed}
Once the present-day values of $H_0$, $w_0$ and $\Omega_{\phi,0}$ are fixed, $w_a$ depends only on the slope of the potential at $\phi_0$, which can in turn be expressed as
\begin{eqaed}
    \dfrac{V'(\phi_0)}{H_0^2}=-\frac{\sqrt{3\Omega_{\phi,0} } \left(3 \left(1-w_0^2\right)-w_a\right)}{2 \sqrt{1+w_0}}\,.
\end{eqaed}
The DESI results then translate into bounds on this slope,
\begin{equation}
\begin{aligned}
\dfrac{V'(\phi_0)}{H_0^2} &= -2.705 \pm 0.377 \quad (\text{DESI+CMB+PantheonPlus})\,,\\
\dfrac{V'(\phi_0)}{H_0^2} &= -3.435 \pm 0.362 \quad (\text{DESI+CMB+Union3})\,,\\
\dfrac{V'(\phi_0)}{H_0^2} &= -3.125 \pm 0.318 \quad (\text{DESI+CMB+DESY5})\,,
\end{aligned}
\end{equation}
where we have assumed without loss of generality $\dot\phi>0$. The field is therefore rolling down a slope, and the slope must be of the same order as the energy density in Planck units. If, in addition, the potential at large redshift corresponds to a maximum or to an asymptotic plateau, the potential must be concave around $\phi_0$. The corresponding slow-roll parameter is
\begin{eqaed}
    \epsilon=\dfrac{1}{2}\left(\dfrac{V'}{V}\right)^2=\frac{ \left(3 \left(1-w_0^2\right)-w_a\right)^2}{6\,(1-w_0)^2\,(1+w_0)\,\Omega_{\phi,0}}\,,
\end{eqaed}
which is of order one for typical values of $w_0$ and $w_a$, indicating that a slow-roll approximation is not well justified in this regime.

These conditions imply a form of double fine-tuning, since both the potential and its first derivative must satisfy
\begin{equation}
V(\phi_0)\simeq V'(\phi_0)\simeq H_0^2\,,
\end{equation}
in Planck units. This differs from the standard quintessence setups designed to reproduce $\Lambda$CDM, where the corresponding double fine-tuning instead reads
\begin{equation}
V(\phi_0)\simeq H_0^2, \qquad V'(\phi_0)\simeq 0\,.
\end{equation}

In fact, given the assumption that dark energy is described by a smooth scalar field, the relevant question to ask is what DESI data constrain within that assumption. In a quintessence scenario, DESI data are primarily sensitive to whether the scalar field is evolving away from a frozen, $\Lambda$CDM like regime or not, so they are most directly sensitive to the first derivative of the potential today. In particular, the DESI-preferred region is characterized by 
\begin{equation}
w_0>-1, \qquad w_a<0\,,
\end{equation}
so that the equation of state becomes more negative as one goes back in time, i.e. toward higher redshift. Imposing unitarity, for instance by assuming quintessence, this points naturally towards a thawing-like evolution, in which dark energy has started evolving more noticeably only recently, and the second derivative may then play an important role in setting the appropriate initial conditions. If one considers a field rolling away from a local maximum or an asymptotic plateau, then one requires
\begin{equation}
V''<0\,.
\end{equation}
Assuming $\dot\phi>0$, the field rolls downhill and therefore $V'<0$. For the field to begin evolving only around the present epoch, the second derivative cannot be arbitrarily small or arbitrarily large compared to Hubble scales. Indeed, if
\begin{equation}
|V''| \ll H_0^2\,,
\end{equation}
the field remains overdamped and the evolution stays too close to slow roll, so that $w \simeq -1$. If instead
\begin{equation}
|V''| \gg H_0^2\,,
\end{equation}
the field evolves too rapidly and oscillates on cosmological scales, so that the assumption of smooth dark energy no longer holds. The phenomenologically interesting window is therefore
\begin{equation}
V'' \sim -H_0^2\,.
\end{equation}
In this regime, the field can remain frozen for most of cosmic history and only start thawing near the present epoch. This naturally points toward a hilltop regime, with the field initially displaced near a local maximum.

By contrast, potentials with positive second derivative do not as naturally reproduce this thawing behavior. In such cases there is no local maximum that can delay the evolution of the field, and one must tune the initial displacement and velocity more strongly in order to reproduce the evolution suggested by the data. One may still obtain viable quintessence dynamics with
$V''>0$,
but in that case the connection between the DESI-preferred evolution and the local shape of the potential is typically less direct than in the hilltop case.
\section{On the choice of initial conditions}\label{choice}

It would also be interesting to understand whether the  initial conditions for the late-time quintessence models considered above can arise from the interaction with a additional scalar field that also describes inflation. A simple possibility is offered by the two-field $\alpha$-attractor models of \cite{Akrami_2018, Kallosh:2025ijd}, in which inflation is driven along an $\alpha$-attractor direction $\chi$ and, after inflation, $\chi$ relaxes and is stabilized at $\chi=0$, while a second field $\xi$ with a much smaller potential drives the late-time expansion. The late-time quintessence potential is then the effective potential for $\xi$ obtained after integrating out the stabilized inflaton, and the initial conditions for $\xi$ today are inherited from the multi-field dynamics. 

We consider a two-attractor toy model
\begin{align}
\mathcal{L}=\frac{R}{2}-3\beta \frac{\partial \chi^2}{(1-\chi^2)^2}-3\alpha \frac{\partial \xi^2}{(1-\xi^2)^2}- V(\chi,\xi)\,,
\end{align}
with scalar potential
\begin{align}
V(\chi,\xi) = V_{1}\,\chi^2 + V_{\rm int}(\chi,\xi) + V_0\,(\xi-c)^2\,.
\label{eq:Vdisk}
\end{align}
This setup is to be read as a proof of principle that the location of the quintessence field near the knee may be inherited from an earlier multi-field evolution, rather than as a fully unified model of cosmological history. The canonically normalized fields are
\begin{align}
\chi=\tanh\!\Big(\frac{\varphi}{\sqrt{6\beta}}\Big)\,,
\qquad
\xi=\tanh\!\Big(\frac{\phi}{\sqrt{6\alpha}}\Big)\,,
\end{align}
in terms of which the potential reads
\begin{align}
V(\varphi,\phi)=V_{\rm 1}\,\tanh^2\!\Big(\frac{\varphi}{\sqrt{6\beta}}\Big)+V_0\Bigg(\tanh\!\Big(\frac{\phi}{\sqrt{6\alpha}}\Big)-c\Bigg)^2+V_{\rm int}(\varphi,\phi)\,,
\label{eq:Vcan}
\end{align}
where $\varphi$ plays the role of the inflaton and $\phi$ that of the quintessence field.

The interaction term $V_{\rm int}$ is to be chosen such that, during inflation ($\varphi\gg \sqrt{6\beta}$), the potential develops a $\varphi$-dependent minimum $\phi_{track}(\varphi)$ defined by
\begin{align}
\frac{\partial V(\varphi,\phi)}{\partial\phi}\bigg\vert_{\phi=\phi_{track}}=0\,,
\end{align}
with mass
\begin{align}
m_{\phi}^2(\varphi)
=
\frac{\partial^2V(\varphi,\phi)}{\partial\phi^2}\bigg\vert_{\phi=\phi_{track}}\gg H^2\,,
\end{align}
so that $\phi$ tracks the instantaneous minimum, $\phi=\phi_{track}(\varphi)$. We further require that during inflation
\begin{align}
V_1 \gg V_{\rm int}(\varphi,\phi_{track})\,, 
\end{align}
so that the inflationary plateau dominates the energy density, and that around $\varphi=0$
\begin{align}
\partial_{\varphi} V_{\rm int}(\varphi,\phi)\simeq 0\,,
\end{align}
such that the inflaton can oscillate around its minimum, allowing for  reheating. Under these assumptions, during inflation the potential is well approximated by
\begin{align}
V(\varphi)
\simeq
V_{\rm 1}\,\tanh^2\!\Big(\frac{\varphi}{\sqrt{6\beta}}\Big)\,,
\end{align}
while the quintessence direction remains stabilized along the tracking solution. After inflation, $\varphi$ settles near its minimum and the leading contribution becomes
\begin{align}
V(\phi)
\simeq
V_0\Bigg(\tanh\!\Big(\frac{\phi}{\sqrt{6\alpha}}\Big)-c\Bigg)^2\,.
\end{align}

\begin{figure}
    \centering
\begin{tabular}{cc}
\hspace{-1cm}\includegraphics[width=0.49\linewidth,trim=0.5cm 0cm 1cm 0cm,clip]{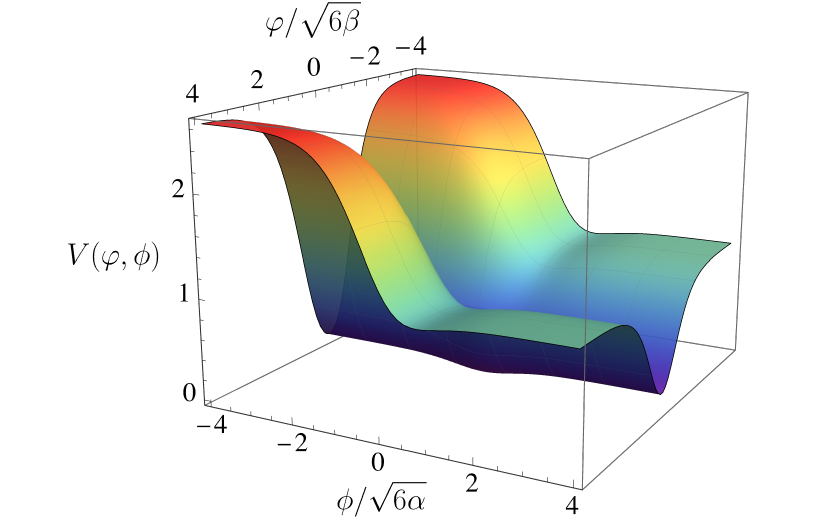}
&
\includegraphics[width=0.40\linewidth]{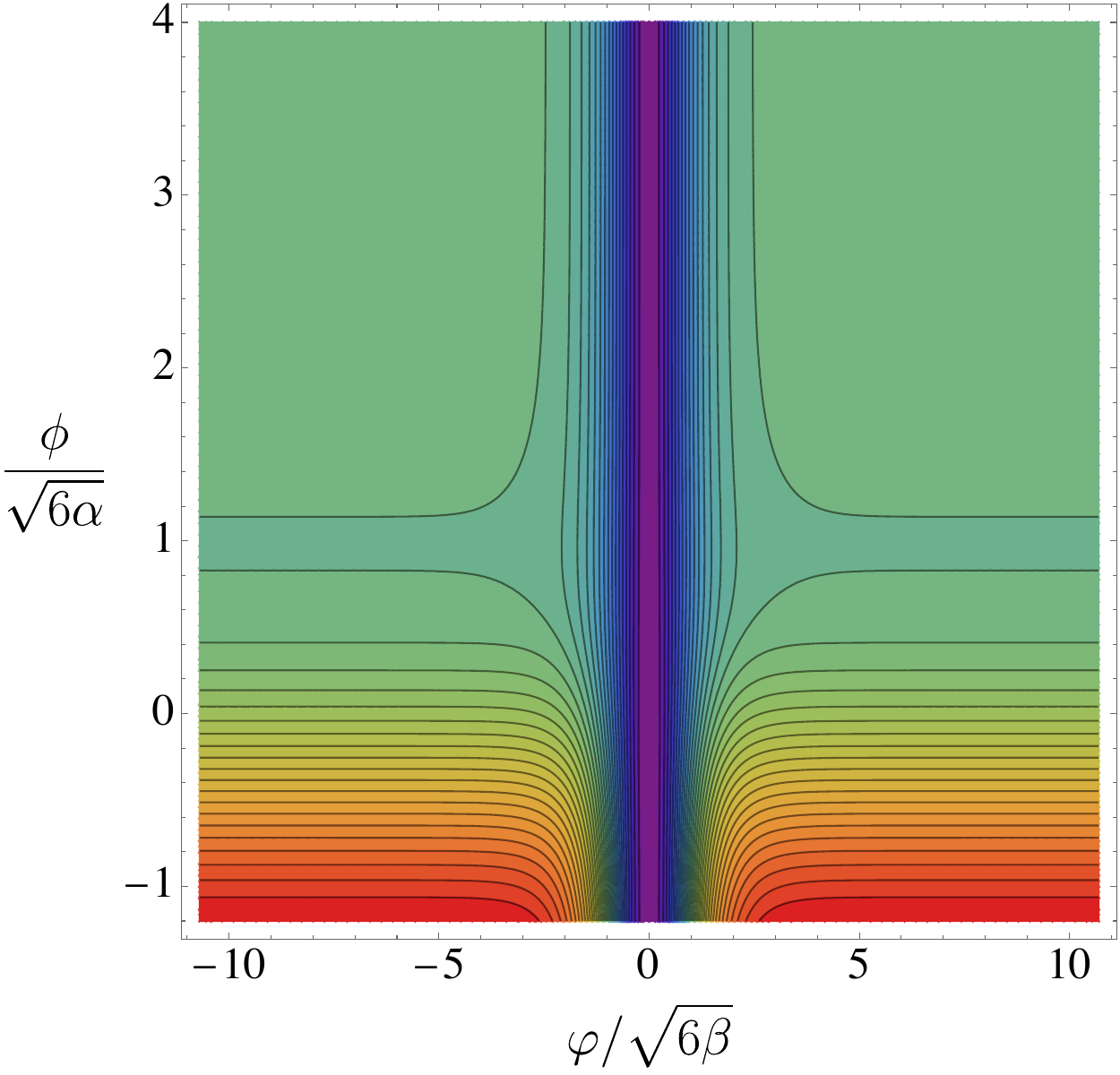}
\end{tabular}
    \caption{\textit{Left: Scalar potential \eqref{eq:Vcan} with $V_{\rm int}$ given by \eqref{eq:Vint}, for $\alpha =\frac{1}{3},\, \beta =\frac{7}{3},\, \lambda =0.5,\, V_0 = 0.1,\, V_1=1,\,c=0$ and $ \,\phi_\star=\sqrt{6 \alpha }$. The inflaton $\varphi$ rolls down its potential, and when inflation ends $\phi$ starts to roll at $\phi\simeq\sqrt{6\alpha}$, at the knee of the late-time potential.  Scales in the plot are just to aid visibility, as realistic scales would require $V_1/V_0\simeq 10^{-110}$.} Right: Contour plot showing the inflationary valley at $\phi\simeq\sqrt{6 \alpha }$. The parameters $\alpha =\frac{2}{3},\, \beta =\frac{7}{3},\, \lambda =0.5,\, V_0 = 0.01,\, V_1=1,\,c=0$ have been chosen to highlight the structure of the potential.}
    \label{fig:3d}
\end{figure}

The $\phi$ initial conditions for quintessence can be estimated by determining when the tracking solution breaks down, namely when
\begin{align}
\frac{\partial^2V(\varphi_{*},\phi)}{\partial\phi^2}\bigg\vert_{\phi=\phi_{track}}
\simeq H^2\,.
\end{align}
At this point $\phi$ freezes out and its evolution becomes dominated by Hubble friction.

For a viable quintessence scenario, $\phi$ must begin to roll around the knee of the potential at the time when the matter density becomes comparable to the dark energy density. This requires the tracking solution to end near the end of the plateau,
\begin{align}
\phi_{track}(\varphi_*)\simeq \sqrt{6\alpha}\,.
\end{align}
This follows from the asymptotic relation $\tanh x \simeq 1 - 2 e^{-2x}$ for large $x$, which shows that the plateau regime breaks down when $x\sim \mathcal O(1)$. The condition translates into a constraint on $V_{\rm int}$, which must be such that the breakdown of the tracking solution occurs at field values corresponding to the knee of the quintessence potential.
An example of an interaction term that leads to the right initial conditions is a dimension-6 term
\begin{equation*}
    V_{\text{int}}=\lambda\, \chi^4 \left(\xi-\xi_\star\right)^2\,,
\end{equation*}
\begin{align} \label{eq:Vint}
    V_{\text{int}}
    =\lambda  \tanh ^4\left(\frac{\varphi }{\sqrt{6\beta} }\right)
    \left(\tanh \left(\frac{\phi }{\sqrt{6\alpha}}\right)-\tanh \left(\frac{\phi_\star}{\sqrt{6\alpha}}\right)\right)^2\,,
\end{align}
where 
$\phi_\star=\sqrt{6 \alpha}\,\text{arctanh}\,\xi_{\star}$ is a constant, the two field potential with this interaction term is shown in Fig. \ref{fig:3d}. Such higher-dimensional couplings are generically expected in an effective field theory approach \cite{Burgess:2007pt}, but here we consider them only as a toy example. We assume that the scales $V_0\gg \lambda\gg V_1$ are such that inflation and quintessence occur largely unaffected by the interaction term. For our toy model the condition $\lambda\gg V_1$ ensures that after inflation $\phi$ freezes at $\phi\simeq \phi_\star$ with $m_{\phi}^2\gg H^2$. Consequently, in order to obtain the right initial conditions it is enough to set
\begin{align}
\phi_{\star}\simeq \sqrt{6 \alpha}\,.
\end{align}
The quintessence field will then track along the inflationary trajectory, and freeze around $\phi\simeq\sqrt{6\alpha}$, in the knee regime.

\section{Discussion}

In this letter, we have studied the relation between quintessential $\alpha$-attractors of the form \eqref{eq:V} and recent DESI indications for dynamical dark energy. Rather than focusing on the asymptotic plateau regime of the attractor potential, we have concentrated on the intermediate field range, namely the \textit{knee}, where the potential interpolates between its asymptotic behaviours and is neither approximately exponential nor quadratic. This region is particularly relevant for thawing quintessence, in which the field remains frozen for most of cosmic history and begins to evolve only near the present epoch.

A least-squares fit between the attractor and the axion-like benchmark potential, performed directly in the field range relevant for quintessence, yields a simple empirical relation between the effective decay constant $f_a$ and the attractor parameter $\alpha$, with a mild dependence on the asymmetry parameter $c$. The DESI constraints on $f_a$ then translate into a preference for $\alpha=\mathcal{O}(1)$, indicating that attractor models in the range $3\alpha = 1,\dots,7$, motivated by supergravity and string constructions, are naturally favored by the data.

To verify this picture dynamically, we solved the cosmological evolution numerically for the attractor potential and computed the corresponding equation of state $w(z)$. The resulting evolution reproduces well the DESI-preferred behaviour over the redshift range relevant to observations.

We argue that the attractor potential can realize a thawing evolution in which the field is initially frozen by Hubble friction and only starts rolling appreciably near the present epoch, similar to the axion potential, and  that the DESI-preferred region can be interpreted within a quintessence framework, as a constraint on the shape of the scalar potential today. Once the present-day values of $H_0$, $\Omega_{\phi,0}$ and $w_0$ are fixed, the parameter $w_a$ is controlled by the first derivative of the potential at the current field position. The preferred region then points toward a situation in which the potential energy and its slope are both of order $H_0^2$ in Planck units. This differs from the near $\Lambda$CDM quintessence regimes, where one instead requires an almost vanishing slope. In this sense, the present observations favor a region of the parameter space in which the field is not exactly frozen today but is already evolving. The second derivative of the potential is also constrained, albeit not as strongly. In particular, the field must remain frozen for most of cosmological history and begin evolving only recently, which naturally favors a concave region of the potential with $V''<0$ and, more specifically, with curvature of order the Hubble scale. This makes hilltop-like or knee-like regions especially suitable, while convex potentials generally require a stronger tuning of the initial conditions in order to reproduce the same behavior. The success of both the axion and attractor potentials can then be interpreted as the fact that they have concave regions with $V\simeq V'M_{Pl}\simeq H^2 M_{Pl}^2$.

Since the attractor potential must be probed near its knee today, it is natural to ask whether such initial conditions can emerge dynamically from an earlier cosmological phase rather than being imposed by hand. As a proof of principle, we discussed a simple two-field attractor setup in which inflation is driven by one field, while a second field later plays the role of quintessence. In this framework, a suitable interaction term can stabilize the quintessence direction during inflation and subsequently release it near the knee of the late-time potential when the tracking solution breaks down. Although the construction we presented is only a toy model and does not yet constitute a complete unified cosmological scenario, it shows that the required late-time initial conditions may arise dynamically in a broader multi-field setting.

Our results suggest that $\alpha$-attractors provide a simple example of quintessence models that can reproduce the  DESI-preferred dark energy evolution, and that the attractor values expected from supergravity/string theory are also within the preferred experimental region. It would be interesting to extend this analysis accounting simultaneously for inflation, reheating and late-time acceleration, which we leave for future work.

\section*{Acknowledgments}
We are thankful to Renata Kallosh and Andrei Linde for stimulating comments on a preliminary version of this work. We also thank David Andriot for stimulating discussions. The work of AB is supported by Regione Sicilia, Avviso 15/2024 (PR FSE+ 2021-2027).  M.S. acknowledges the support of the University of Catania through PIAno di inCEntivi per la RIcerca di Ateneo 2024/2026 - Project "COSMOgraM". 
\printbibliography
\end{document}